\documentclass[final,3p,times]{elsarticle}
\usepackage{ecrc}

\volume{00}

\firstpage{1}

\journalname{Physica C}

\runauth{M. Arivazhagan et al.}

\jid{Physica C}

\jnltitlelogo{Physica C}

\usepackage[dvipsnames]{xcolor}
\usepackage{amssymb,amsmath}

\usepackage[figuresright]{rotating}
\usepackage{multirow}

\begin{document}

\begin{frontmatter}




\title{Parametric triggering of vortices in toroidally trapped rotating Bose-Einstein Condensates}


\author[hhr]{M. Arivazhagan}
\author[bdu]{P. Muruganandam}
\ead{anand@bdu.ac.in}
\author[hhr,cauth]{N. Athavan}
\ead{n.athavan@gmail.com}
\address[hhr]{Department of Physics, H. H. The Rajah's College (Autonomous), Pudukkottai 622001, (Affiliated to Bharathidasan University, Tiruchirapalli), Tamilnadu, India}
\address[bdu]{Department of Physics, Bharathidasan University, Tiruchirappalli 620024, Tamilnadu, India}
\cortext[cauth]{Corresponding author}




\begin{abstract}
We study the creation of vortices by triggering the rotating  Bose-Einstein condensates in a toroidal trap with trap parameters such as laser beam waist and Gaussian potential depth. By numerically solving the time-dependent Gross-Pitaevskii equation in two dimensions, we observe a change in vortex structure and a considerable increase in the number of vortices when the waist of the irradiated laser beam is in consonance with the area of the condensate as we vary the Gaussian potential depth. By computing the root mean square radius of the condensate, we confirm the variation in the number of vortices generated as a function of the ratio between the root-mean-square radius of the condensate and the laser beam waist. In particular, the number of hidden vortices reaches the maximum value when the above ratio is close to the value 0.7. We find the variation in the number of vortices is rapid for deeper Gaussian potentials, and we conclude that the larger beam waist and deeper Gaussian potentials generate more vortices. Further, we calculate the number of vortices using the Feynman rule with Thomas Fermi approximation and compare them with the numerical results. We also observe that the critical rotation frequency decreases with an increase in depth of Gaussian potential.
\end{abstract}


\end{frontmatter}



\section{Introduction}
\label{sec:intro}
A most intriguing occurrence in the realm of rotating Bose-Einstein condensates (BECs) is the creation of quantized vortices like superfluidity of liquid helium~\cite{Onsager1949}. Quantum vortices are nonlinear physical entities that arise due to the excitation of  BECs~\cite{Madison2000, AboShaeer2001, Matthews1999, Raman2001, Dutton2001, Anderson2001}. Experimentally vortices in BEC have been generated by either applying rotating magnetic traps or laser stirring, and they are usually formed above a critical rotation frequency. In a BEC, the vortices have quantized circulation, and their interaction on inhomogeneous density background have been studied extensively~\cite{Mason2008, Nilsen2006, Li2008} complementing the theoretical investigations~\cite{Pismen1999, Fetter2001} over the past few decades.

On the other hand, the observation of BEC in a toroidal geometry has opened up a new window for remarkable studies about their fundamental properties including persistent currents~\cite{Javanainen1998, Dubessy2012}, ring dark solitons~\cite{Toikka2012, Toikka2013, Toikka2014}, weak links~\cite{Ramanathan2011, Wright2013, Piazza2013}, self interference~\cite{Toikka2014a}, symmetry breaking~\cite{Oliinyk2019}, atomic-phase interference ~\cite{Anderson2003}, decay of persistent current via phase slips~\cite{Moulder2012}, creation of a vortex-antivortex pair, multiple quantized vortices~\cite{Cozzini2006, Fischer2003, Jackson2004, Aftalion2004}, hidden vortices~\cite{Wen2010, Wen2012}, transition from vortex excitation to solid-body like motion \cite{Oegren2021}, transition from quantum anti-resonance to dynamical localization \cite{Wang2021}, half-quantum vortex and complex vortex lattice \cite{Yang2019}. Toroidally trapped Bose-Einstein condensates have been experimentally realized by superimposing a laser beam on a harmonically trapped condensate, thus creating a harmonic plus Gaussian potential~\cite{Bretin2004, Stock2004, Ryu2007, Weiler2008}, which also favours the occurrence of vortices under suitable conditions.

By using a toroidal trap geometry, one can show that the formed condensate can either be like a disc or an annulus depending upon the parameters of the system, such as rotational frequency ($\Omega$), intensity ($V_0$) and waist of Gaussian potential depth of the laser beam~\cite{Aftalion2010}. To be precise, the condensate bears a disc-like structure when the rotational frequency ($\Omega$) is notably smaller than radial trap frequency ($\omega_\perp$), and on the other hand, rapid rotation ($\Omega/\omega_\perp \sim 1$) makes the condensate to expand resulting in a large annulus with a vortex lattice inside the condensate, and a large circulation within the central hole~\cite{Mason2009, Fetter2009}. Moreover, the vortex dynamics of rotational BECs trapped by harmonic plus Gaussian potential for varied rotational velocity has been recently analysed in reference \cite{Aftalion2010}. However, they have investigated the dynamics of the vortices for different rotational frequencies by keeping the beam waist constant. In contrast to these above investigations, we intend to study the vortex dynamics of BECs in such a toroidal trap as a function of trap parameters, namely the potential depth and waist of Gaussian potential under constant rotation. In particular, by altering the trap parameters suitably, we explore the possibility of creating more number of new vortices in the rotating BECs and the maximum extent of vortex number for a particular choice of beam waist. Additionally, we analyze the relationship between the critical rotational frequency and the depth of the Gaussian potential.

By numerically solving the quasi-two-dimensional Gross-Pitaevskii equation, we calculate the number of vortices as a function of the potential (Gaussian) depth for different waist lengths of Gaussian laser, and compare them with that of calculated values by adhering to the Feynmann rule with the Thomas-Fermi (TF) approximation. We have found that the increase in Gaussian potential depth hastens the vortex formation in BECs. Also, we have observed that the critical rotation frequency decreases when the depth of the Gaussian potential increases for the same beam waist. We have calculated other associated quantities, such as the expectation value of angular momentum, root mean square (RMS) radius, and chemical potential of the BEC as a function of Gaussian potential depth ($V_0$) separately for three different beam waist values.

This paper is organised as follows. In section~\ref{sec:frame}, we provide the theoretical background by describing the dynamics of rotating BEC confined in an axially symmetric toroidal trap with the aid of the mean-field Gross-Pitaevskii equation. Then, in section~\ref{sec:results}, we numerically illustrate the formation of vortices and their dynamical behaviour with respect to the change in the quantities which shape the Gaussian potential. We also investigate the role played by the waist of the Gaussian laser beam on the enhancement of the number of vortices. Finally, in section~\ref{sec:conclu} we provide a summary and conclusion.

\section{Theoretical background}
\label{sec:frame}
\subsection{Evolution Equation}
According to the mean-field approximation, the dynamics of a rotating BEC with $N$ atoms, each of mass $m$ at an absolute zero temperature in a rotating frame can be described by the dimensionless form of the 2D Gross-Pitaevskii (GP) equation, which reads as~\cite{Adhikari2002, Bao2006}
\begin{align} 
i\frac{\partial\phi(r,t)}{\partial t}  = \left[-\frac{1}{2}\nabla^2 +V(r)+ g \vert\phi(r,t)\vert^2 -\Omega L_z \right]\phi(r,t), \label{eqn:gp}
\end{align}
with $i=\sqrt{(-1)}$. Here $V(r)$, $r \equiv(x,y)$, is the external trapping potential, $g = 4\pi a N/\sqrt{2\pi} d_z$ is a two-dimensional coupling parameter, $\phi(r,t)$ is the wave function at time $t$, and $a$ is atomic scattering length which can be tuned via Feshbach resonance. $L_z = - i (x\partial_y-y \partial_x)$ corresponds to the $z$-component of the angular momentum due to the rotation of the BEC about $z$ axis with angular velocity $\Omega$. The radial and axial trap frequencies of the harmonic potential, $V_{ha}$, are $\omega_\perp$ and $\omega_z$, respectively, and are related with the trap aspect ratio, $\lambda \equiv \omega_z/\omega_\perp$ and $d_z = 1/\sqrt{\lambda}$. It is to be noted that in Eq.~(\ref{eqn:gp}) length is measured in units of characteristic harmonic oscillator length $\sqrt{\hbar/m\omega_\perp}$, frequency in units of $\omega_\perp$ and time in units of $\omega_\perp^{-1}$.

The normalization condition of the wave function is
\begin{align}
\int \vert\phi(r,t)\vert^2 d\tau = 1, \label{eqn:norm}
\end{align}
where $d\tau$ is the spatial integral element and the chemical potential in this case for the state $\phi(r,t)$ can be written as
\begin{align}
\mu = \int \left[\frac{1}{2} \vert\nabla\phi\vert^2+V(r)\vert\phi\vert^2+g \vert\phi\vert^4-{\phi^*}\Omega L_z\phi\right]d\tau, \label{eqn:chem}
\end{align}
where ${\phi^*}$ is the complex conjugate of the wave function $\phi$ and $L_z$ is the $z$ component of the angular momentum.

The dimensionless 2D toroidal trap potential $V(r)$ is given by
\begin{align}
V(r) = \frac{r^2}{2} + V_0 \exp(-2r^2/w_0^2), \label{poten}
\end{align}
with Gaussian potential depth $V_0$ with units of $\hbar\omega_\perp$, and $w_0$ is the waist of laser beam in units of $\sqrt{\hbar/m\omega_\perp}$.

\subsection{Root mean square radius of the condensate}

We determine the rms radius ($r_{rms}$) with respect to $V_0$ in the mean-field TF regime. When the interaction energy is larger than the kinetic energy, the kinetic energy can be neglected and in doing so we enter into the TF regime in Eq.~(\ref{eqn:gp}) and therefore the approximation yields the expression for the TF density as~\cite{Aftalion2010}
\begin{align}
g\vert\phi_{TF}\vert^2=\mu_{TF}-\frac{1}{2}(1-\Omega^2)r^2-V_0 e^{-\frac{2r^2}{w_0^2}}, \label{eqn:TFden}
\end{align}
where $\mu_{TF}$ is the chemical potential in the TF regime.

As a next step we deduce $r_{rms}$ for the two limiting values of laser waist $w_0$. When $w_0\gg r$, or in other words, the laser beam waist is comparatively larger than the outer TF radius of the condensate, $2r^2/w_0^2$ is very small, that is $(\ll 1)$, and hence the expansion of the exponential term in the expression for the TF density in Eq.~(\ref{eqn:TFden}) yields
\begin{align}
g\vert\phi_{TF}\vert^2=\mu_{TF}^{'}-\frac{1}{2}\left(1-\Omega^2-\frac{4V_0}{w_0^2}\right)r^2, \label{eqn:TFdensmall}
\end{align}
where $\mu_{TF}^{'}=\mu_{TF}-V_0$.

When $4V_0/w_0^2$ becomes larger than $1-\Omega^2$, we consider the next higher order term of the exponential function in Eq.~(\ref{eqn:TFden}).

It may be noted that the density of the condensate $\vert\phi_{TF}\vert^2$ vanishes at radius $R_{TF}$, which in turn leads to the following chemical potential relation with frequency of rotation ($\Omega$) from Eq.~(\ref{eqn:TFdensmall}), 
\begin{align}
\mu_{TF}^{'}(\Omega)=\frac{1}{2}\left(1-\Omega^2-\frac{4V_0}{w_0^2}\right)R_{TF}^2(\Omega). \label{eqn:TFchem1}
\end{align}
By using the above Eq.~(\ref{eqn:TFchem1}), we get the ratio of square of the TF radius with and without rotation as
\begin{align}
\frac{R^2_{TF}(\Omega)}{R^2_{TF}(0)}=\frac{\mu_{TF}^{'}(\Omega)}{\mu_{TF}^{'}(0)}\left(\frac{1-\Omega^2-\frac{4V_0}{w_0^2}}{1-\frac{4V_0}{w_0^2}}\right)^{-1}. \label{eqn:radratio}
\end{align}
After substituting the expression for $\vert\phi_{TF}\vert^2$ in Eq.~(\ref{eqn:TFdensmall}) into the normalization condition in Eq.~(\ref{eqn:norm}) and integrating it over the domain between $0$ and $R_{TF}$, we get
\begin{align}
\mu_{TF}^{'}(\Omega)-\frac{1}{4}\left(1-\Omega^2-\frac{4V_0}{w_0^2}\right)R_{TF}^2(\Omega)
= \frac{g}{\pi R_{TF}^2(\Omega)}. \label{eqn:TFrelation}
\end{align}
We now substitute for $R_{TF}^2$ from Eq.~(\ref{eqn:TFchem1}) in Eq.~(\ref{eqn:TFrelation}) and get 
\begin{align}
\mu_{TF}^{'}{}^2(\Omega)=\frac{g}{\pi}\left(1-\Omega^2-\frac{4V_0}{w_0^2}\right). \label{eqn:chem1}
\end{align}
Then the ratio of chemical potentials with and without rotation can be written as
\begin{align}
\frac{\mu_{TF}^{'}(\Omega)}{\mu_{TF}^{'}(0)}=\left(\frac{1-\Omega^2-\frac{4V_0}{w_0^2}}{1-\frac{4V_0}{w_0^2}}\right)^{\frac{1}{2}}. \label{eqn:chemratio}
\end{align}
Now putting Eq.~(\ref{eqn:chemratio}) in Eq.~(\ref{eqn:radratio}), we get the TF radius as a function of rotation ($\Omega$) as
\begin{align}
R_{TF}(\Omega)=R_{TF}(0)\left(\frac{1-\Omega^2-\frac{4V_0}{w_0^2}}{1-\frac{4V_0}{w_0^2}}\right)^{-\frac{1}{4}}. \label{eqn:TFrad}
\end{align}
In order to get an expression for $R_{TF}(0)$, we equate Eqs.~(\ref{eqn:TFchem1}) and ~(\ref{eqn:chem1}) after substituting $\Omega=0$ and obtain $R_{TF}(0) = \left[ 4g/\pi(1 - \frac{4V_0}{w_0^2}) \right]^{\frac{1}{4}}$. By using the same analysis, one can obtain the expression for the outer radius of the condensate $(R_2)$ and subsequently the inner radius $(R_1)$ after eliminating the higher order terms in Eq.~(\ref{eqn:A1}). If $w_0\ll r$, that is the laser beam waist is relatively small compared to the outer TF radius of the condensate, $2r^2/w_0^2$ becomes very large, that is $(\gg 1)$, and the exponential term in Eq.~(\ref{eqn:TFden}) can be neglected. Finally, we get radius of the condensate as
\begin{align}
R_{TF}(\Omega)= R_{TF}(0)(1-\Omega^2)^{-1/4}, \label{eqn:TFradlarge}
\end{align}
where $R_{TF}(0)=(4g/\pi)^{1/4}$. One can easily obtain the expressions for outer and inner radii of the condensate  using the expressions given in Eqs.~(\ref{eqn:A4}) and ~(\ref{eqn:A5}) of Appendix A, respectively. $R_{TF}(0)$ is related to  root mean square radius $r_{rms}$ by $R_{TF}(0) = \sqrt{3}r_{rms}$~\cite{Abad2010,kishor2012}. 

\subsection{Vortex Number}

The dimensionless form of Feynman rule under equilibrium for the number of vortices in a superfluid rotating in a rigid container is given by $N_v=\Omega R_\perp^2(\Omega)$, where $R_\perp(\Omega)$ is the radius of the rotating fluid~\cite{Feynman1955}. As one needs a radius value for the numerical estimation of the number of vortices, root-mean-square radius value may be conveniently chosen for obtaining the number of vortices closer to the values suggested by Feynman rule as in ref.\cite{kishor2012}.
 
Substituting the expression for $R_{TF}^{2}(\Omega)$ from Eqs.~(\ref{eqn:TFrad}) and (\ref{eqn:TFradlarge}) the Feynman's relation (number of vortices) for the case of rotating BEC trapped under harmonic and Gaussian potential with large laser waist and with small laser waist are
\begin{align}
N_v=\Omega R_\perp^2(0)\left(\frac{1-\Omega^2-\frac{4V_0}{w_0^2}}{1-\frac{4V_0}{w_0^2}}\right)^{-1/2},	\label{eqn:vnosmall}
\end{align}
and 
\begin{align}
N_v=\Omega R_\perp^2(0)(1-\Omega^2)^{-1/2}, \label{eqn:vnolarge}
\end{align}
respectively.

\subsection{Expectation value of Angular momentum}

The birth of the new vortices can be identified by calculating the  increase in the value of average angular momentum per atom ($\langle L_z\rangle$) as we increase the Gaussian potential depth ($V_0$).   $\langle L_z\rangle$ is calculated by using the following expression
\begin{align}
\left\langle L_{z}\right\rangle=i \int \phi^{*}(r, t)\left(y \partial_{x}-x \partial_{y}\right) \phi(r, t)\, d\tau.
\end{align}

\section{Numerical illustration of vortex dynamics}
\label{sec:results}
To numerically illustrate the vortex dynamics of BECs under toroidal trap which mimics the real situation, we consider the respective dimensionless equivalent numerical values of the parameters used by the authors in reference \cite{Bretin2004} during their experimental investigations on  the rotating properties of Bose-Einstein condensate of  ${}^{87}Rb$ atoms confined in toroidal geometry with a weak laser beam. The characteristic harmonic oscillator length of ${}^{87}Rb$ atoms is $1.24\mu m$ and the trap aspect ratio of the condensate used in their experiment was equivalent to $\lambda=0.146$. Precisely, they have used a laser power of $1.2$mW with waist $w_0=25\mu m$ from which one can obtain the dimensionless values for the Gaussian potential depth $V_0$  and the laser waist  $w_0$ as  $24.8$ and $20.2$, respectively. Further, we consider $\Omega=0.5$ to match the experimental value of  37.75Hz and the two-dimensional coupling parameter $g\sim 500$ for $60000$ $^{87}Rb$ atoms. All our numerical simulations have been carried out with $dx=dy=0.2$ (space step) and $dt=0.005$ (time step).

\subsection{Effect of varying the depth and waist of Gaussian potential on the shape of the condensate and number of vortices}

\begin{figure*}[!ht]
\begin{center}
\includegraphics[width=0.9\linewidth]{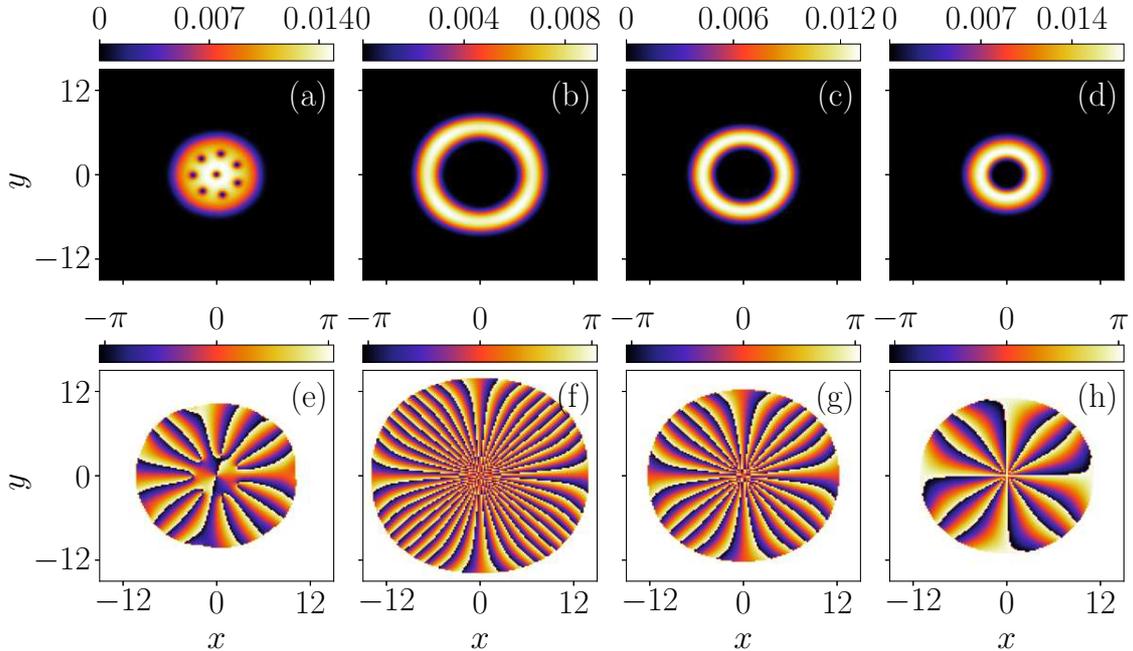}
\end{center}
\caption{Contour plots of the density distributions $\vert\phi\vert^2$ showing the steady vortex state in a rotating BEC under toroidal trap with $g=500$ and $\Omega=0.5$. For $V_0=50$, (a) $w_0=20.2$, (b) $w_0=9.43$, (c) $w_0=4.71$ and (d) $w_0=2.36$. (e)-(h) are the corresponding phase distributions of (a)-(d).}
\label{fig:waist}
\end{figure*}

To start with, we obtain the ground state wavefunction by solving Eq.~(\ref{eqn:gp}) using imaginary time propagation in the presence of harmonic and Gaussian potential without rotation $(\Omega=0)$. Then, the obtained ground state wave function has been allowed to evolve in real time with constant rotation (frequency of rotation $\Omega=0.5$). The dissipation is introduced by replacing ‘$i$’ with ‘($i-\gamma$)’ in the time dependent equation ~(\ref{eqn:gp}) to achieve the equilibrium state, where $\gamma$ ($\sim 10^{-5}$) accounts for the dissipation strength. We have fixed the coupling parameter $g=500$.  The shape of the condensate can be tailor made by choosing the appropriate values for the intensity and the waist of the Gaussian beam such that one can get either a disc or an annular shaped condensate. At first instance, we have investigated the effect on the number of vortices $N_v$ for various laser waist values $w_0=20.2$, 9.43, 4.71 and 2.36 with Gaussian potential depth fixed at $V_0=50$. When we take the largest laser waist $w_0=20.2$, we obtain the corresponding rms radius of the condensate ($r_{rms}$) as 3.83 and the density of the condensate is maximum at the centre (disk) with eight vortices, as shown in Fig.~\ref{fig:waist}(a), and the corresponding phase plot has been shown in Fig.~\ref{fig:waist}(e). %
Next, when we reduce $w_0$ to $9.43$, the rms radius ($r_{rms}$) increases to $6.96$ and the condensate becomes annular with no visible vortices ($N_{vv}$) as shown in Fig.~\ref{fig:waist}(b). However, with the aid of the phase plot, we identify the presence of $20$ hidden vortices ($N_{hv}$) which can be seen from Fig.~\ref{fig:waist}(f). Further reduction in $w_0$ to $4.71$ and then to $2.36$ resulted in the decline of the rms radius ($r_{rms}$) to $5.48$ and $3.98$, respectively. By using the density plot Figs.~\ref{fig:waist}(c)-\ref{fig:waist}(d), we show that the size of the annulus got decreased, and the corresponding phase plot Figs.~\ref{fig:waist}(g)-\ref{fig:waist}(h) show that the number of hidden vortices reduces to $12$ and $6$, respectively. %
\begin{figure}[!ht]
\begin{center}
\includegraphics[width=0.5\linewidth]{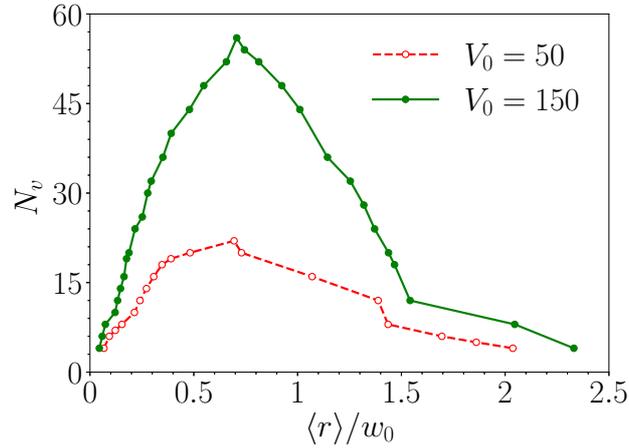}
\end{center}
\caption{Plot of the number of vortices as a function of the ratio of rms radius of condensate and waist of Gaussian beam ($r_{rms}/w_0$) for $V_0 = 50$ and 150. Red colour curves indicate the visible vortices ($N_{vv}$) and green colour curves represent the hidden vortices ($N_{hv}$).}
\label{fig:ratio}
\end{figure}%
The plot, for the number of vortices $(N_v=N_{vv}+N_{hv})$ with respect to the ratio between the rms radius of condensate ($ r_{rms}$) and the waist of Gaussian beam ($w_0$), in Fig.~\ref{fig:ratio} clearly shows the number of hidden vortices ($N_{hv}$) is maximum when $r_{rms}/w_0$  is close to value of 0.7 and $N_v$ decreases as $r_{rms}/w_0$ deviates from that value. Interestingly, from Fig.~\ref{fig:ratio} one can observe an abrupt variation in the number of vortices for deeper Gaussian potential.

Next, we investigate the effect on the number of vortices $N_v$ for a select set of Gaussian potential depths in multiples of 30, say, $V_0=30$, $60$, $90$ and $120$ with different Gaussian beam waists, for instance, $w_0=20.2$, $4.71$ and $0.47$. %
\begin{figure*}[!ht]
\begin{center}
\includegraphics[width=0.9\linewidth]{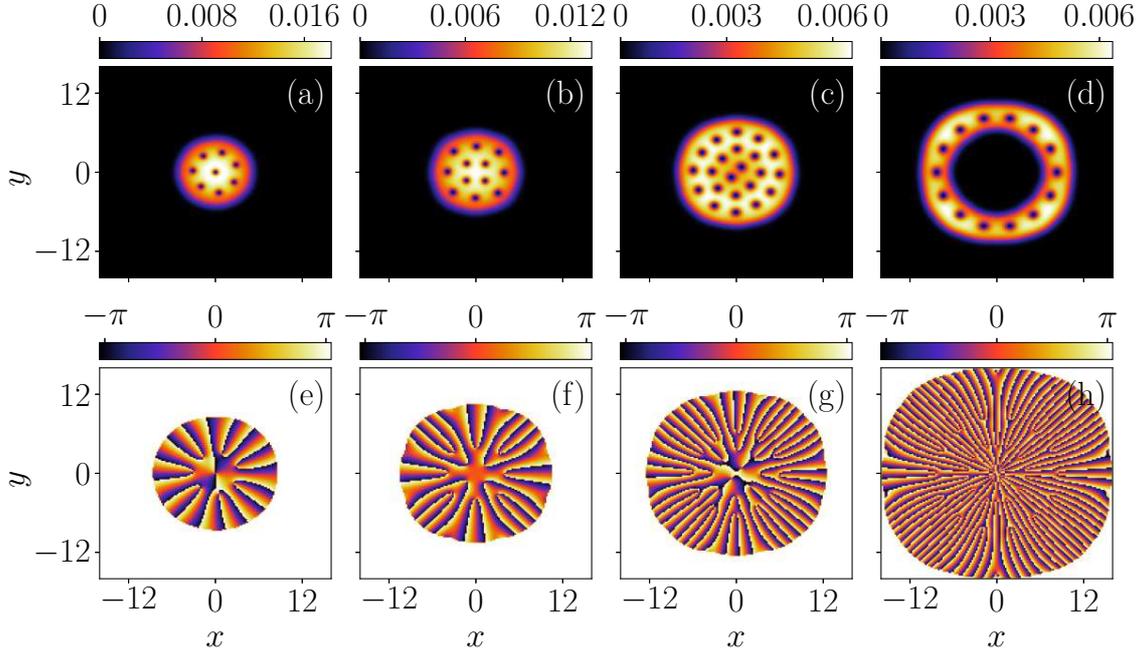}
\end{center}
\caption{Contour plots of the density distributions $\vert\phi\vert^2$ showing the steady vortex state in a rotating BEC under toroidal trap with $g=500$ and $\Omega=0.5$. For $w_0=20.2$, (a) $V_0=30$, (b) $V_0=60$, (c) $V_0=90$ and (d) $V_0=120$. (e)-(h) are the corresponding phase distributions of (a)-(d)}.
\label{fig:consmall}
\end{figure*}
In the case of the largest laser waist $w_0=20.2$, we have seen that the density of the condensate is maximum at the centre for Gaussian depths $V_0=30$  and $V_0=60$ and the corresponding number of vortices in the condensate are eight and twelve as given in Figs.~\ref{fig:consmall}(a) and \ref{fig:consmall}(b), respectively. Figs.~\ref{fig:consmall}(e) and \ref{fig:consmall}(f) show the respective phase plots of Figs.~\ref{fig:consmall}(a) and \ref{fig:consmall}(b). When we increase the Gaussian depth to $V_0=90$, the number of vortices increases to $22$ with local minimum density at the centre of condensate as shown in Fig.~\ref{fig:consmall}(c). Fig.~\ref{fig:consmall}(g) shows the respective phase pattern of Fig.~\ref{fig:consmall}(c). Finally,  when we increase the Gaussian depth to $V_0=120$, the condensate takes up the annular shape with the emergence of 14 visible vortices along the ring as shown in Fig.~\ref{fig:consmall}(d). In fact, we have observed, by using the phase plot Fig.~\ref{fig:consmall}(h),  $24$ hidden vortices in the annulus which are clustered together, forming a large hole at the centre.

\begin{figure*}[!ht]
\begin{center}
\includegraphics[width=0.9\linewidth]{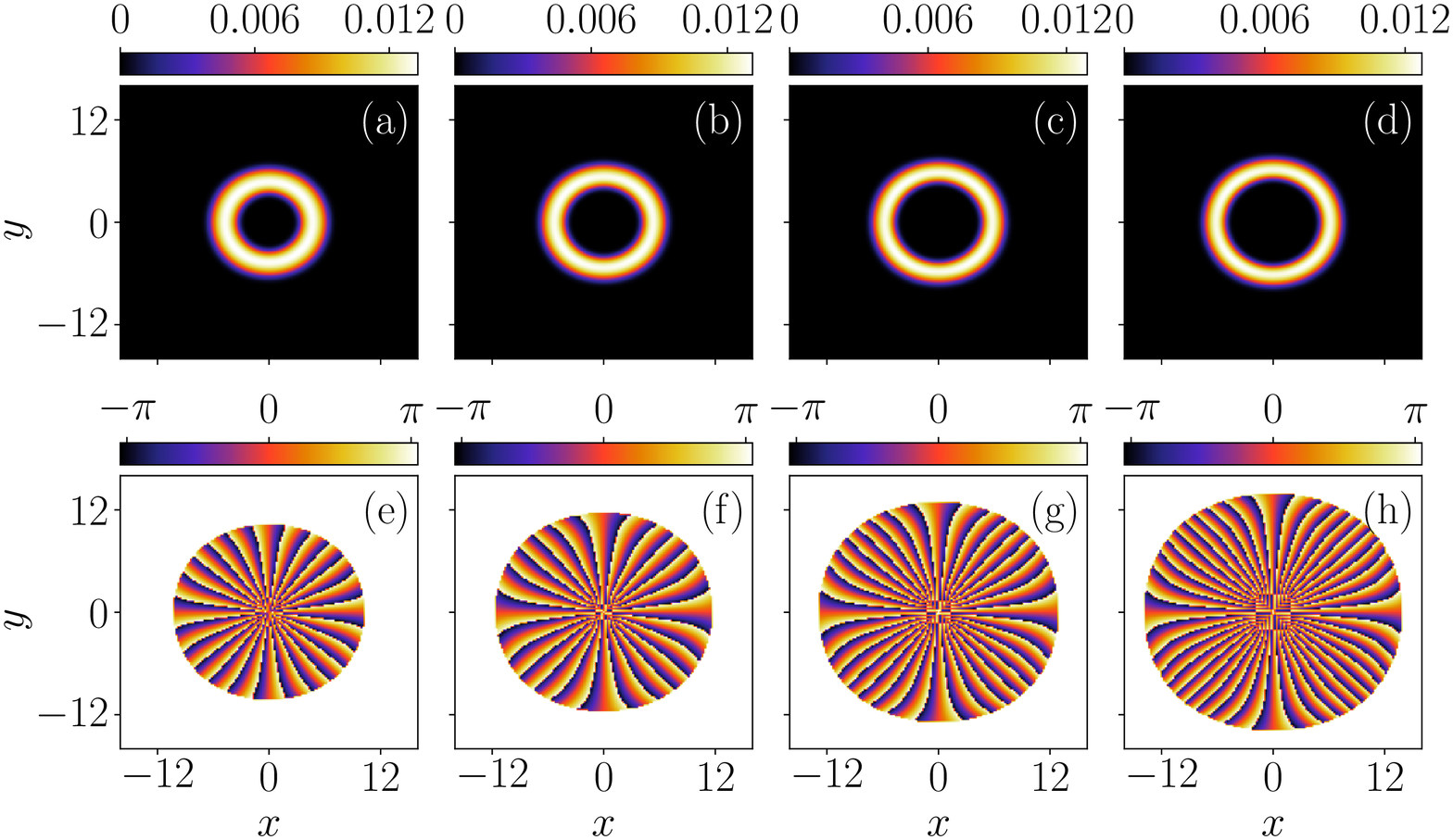}
\end{center}
\caption{Contour plots of the density distributions $\vert\phi\vert^2$ showing the steady vortex state in a rotating BEC under toroidal trap with $g=500$ and $\Omega=0.5$. For $w_0=4.71$, (a) $V_0=30$, (b) $V_0=60$, (c) $V_0=90$ and (d) $V_0=120$. (e)-(h) are the corresponding phase distributions of (a)-(d).}
\label{fig:medium}
\end{figure*}

Now, by considering the intermediate value of laser waist, that is, $w_0=4.71$, we note that, the condensate becomes annular and vortices are hard to see in the condensate as exposed in Figs.~\ref{fig:medium}(a)-\ref{fig:medium}(d). %
For $V_0=30$ and $60$, we identify $12$ hidden vortices from the phase plot Figs.~\ref{fig:medium}(e)-\ref{fig:medium}(f). When $V_0=90$ and $120$, we witnessed more number of hidden vortices (about $16$ to $20$ ) by analysing the corresponding phase plots shown in Figs.~\ref{fig:medium}(g)-\ref{fig:medium}(h).

\begin{figure*}[!ht]
\begin{center}
\includegraphics[width=0.9\linewidth]{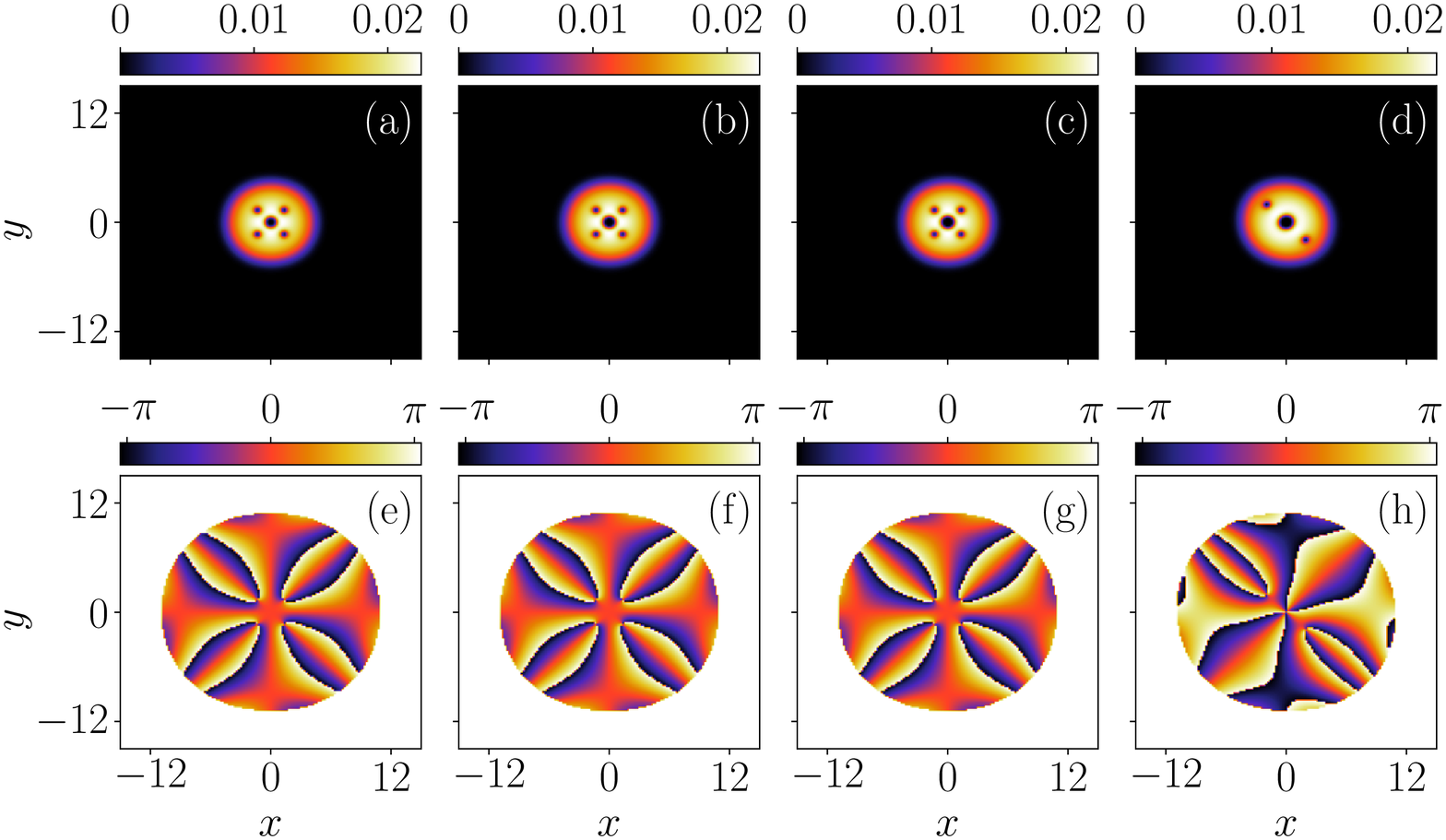}
\end{center}
\caption{Contour plots of the density distributions $\vert\phi\vert^2$ showing the steady vortex state in a rotating BEC under toroidal trap with $g=500$ and $\Omega=0.5$. For $w_0=0.47$, (a) $V_0=30$, (b) $V_0=60$, (c) $V_0=90$ and (d) $V_0=120$. (e)-(h) are the respective phase distributions of (a)-(d)}.
\label{fig:conlarge}
\end{figure*}

By considering the smallest laser waist of Gaussian laser beam $w_0=0.47$ to suit the conditions for condensate formation, we plot the density plots \ref{fig:conlarge}(a)-\ref{fig:conlarge}(c), which show zero density at the centre forming thin annular having four visible vortices that appear in the condensate for $V_0=30$, $60$ and $90$, two visible and two hidden vorices are found in \ref{fig:conlarge}(d) for $V_0=120$. However the total number of vortices in the condensate are invariable in all four values. Figs.~\ref{fig:conlarge}(e)-\ref{fig:conlarge}(h) show the respective phase plots of Figs.~\ref{fig:conlarge}(a)-\ref{fig:conlarge}(d).

From the three sets of the laser waist values ($w_0=20.2$, $4.71$ and $0.47$) along with the corresponding four sets of Gaussian potential depths ($V_0=30$, $60$, $90$ and $120$), we conclude that, the number of vortices is larger for the intermediate laser waist, that is for $w_0=4.71$. However, when the Gaussian potential is hiked from the minimum value, say $V_0=30$, the growth rate of the number of vortices is larger for the larger laser waist, that is for $w_0=20.2$. This is because when Gaussian potential depth ($V_0$) increases, the rms radius of the condensates approaches the value of the waist of the laser beam corresponding to $w_0=20.2$. For $w_0=4.71$, the rms radius value exceeds the value of the waist resulted in poor overlapping. Finally, when we take $w_0=0.47$, rms radius is still larger than the waist which means that a pinpoint like-overlap of Gaussian beam on the condensate takes place. %
\begin{table*}[!ht]
\caption{The comparison of vortex numbers with respect to Gaussian potential depth ($V_0$)  for different laser waists ($w_0$)} \label{tab:results}
	\begin{center}
		\begin{tabular}{r | r r r r r}
			\hline
			\multicolumn{1}{c|}{\multirow{2}{*}{$w_0$}} & \multicolumn{5}{c}{$N_v$} \\
			& \multicolumn{1}{c}{$V_0=30$} & \multicolumn{1}{c}{$V_0=60$} & \multicolumn{1}{c}{$V_0=90$} & \multicolumn{1}{c}{$V_0=120$} & \multicolumn{1}{c}{$V_0=150$} \\
			\hline
			20.20 & 8 & 12 & 22 & 38 & 48  \\
			4.71 & 12 & 12 & 16 & 20 & 24  \\
			0.47 & 4 & 4 & 4 & 4 & 4  \\
			\hline
		\end{tabular}
	\end{center}
\end{table*}%
Hence, even for a larger Gaussian potential depth, we could not find any elevation vortex number. The comparison between the number of vortices for different choices of waist and potential depth are presented in table \ref{tab:results}.

\subsection{Root mean square radius and  chemical potential}

\begin{figure}[!ht]
\begin{center}
\includegraphics[width=0.5\linewidth]{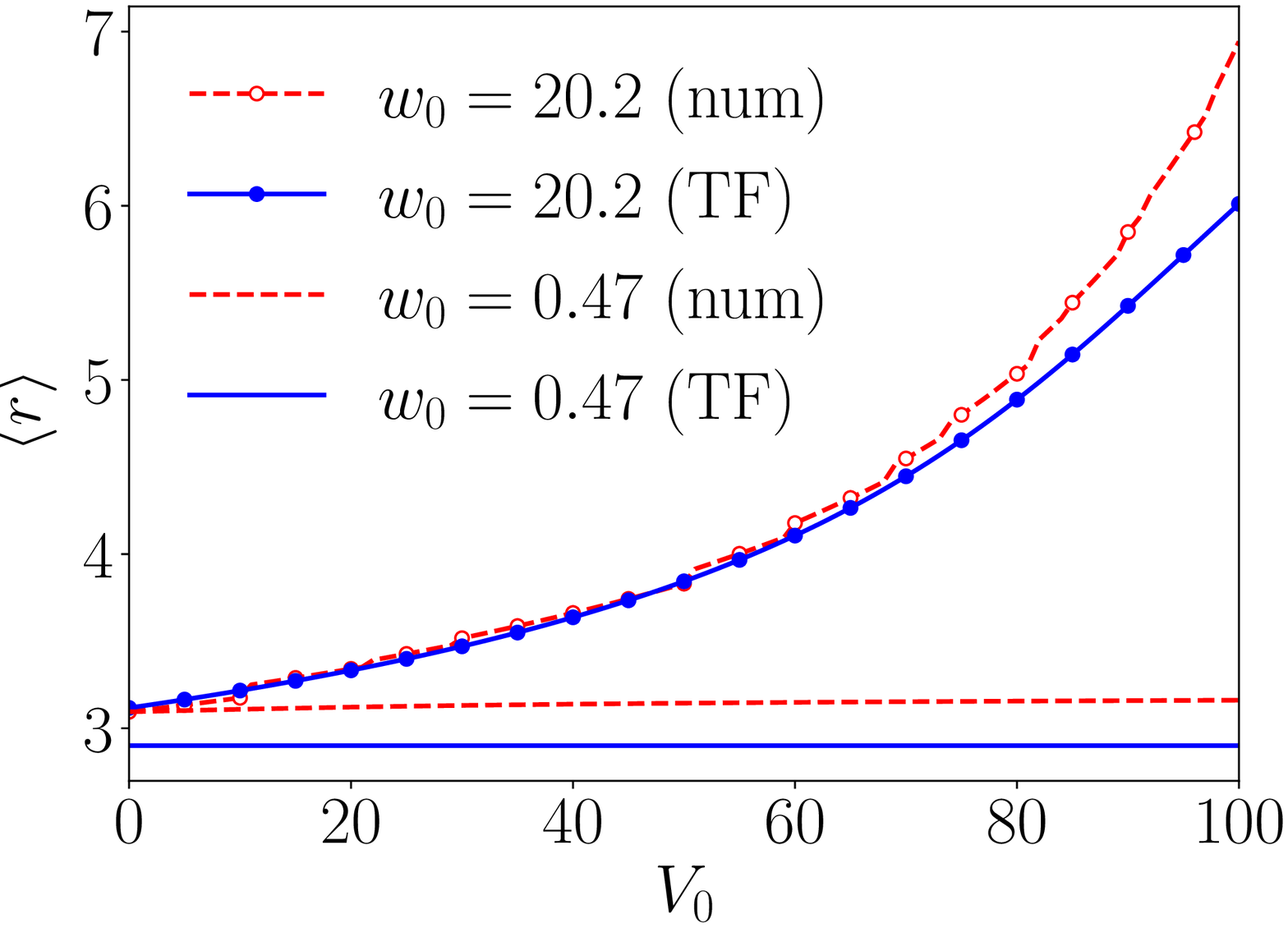}
\end{center}
\caption{Plot of rms radius ($r_{rms}$) as a function of Gaussian potential depth ($V_0$) for $w_0=20.2$ and $w_0=0.47$. Red colour curves represent the numerically calculated values and blue colour curves denote the TF approximated values.}
\label{fig:rmscom}
\end{figure}

\begin{figure}[!ht]
\begin{center}
\includegraphics[width=0.5\linewidth]{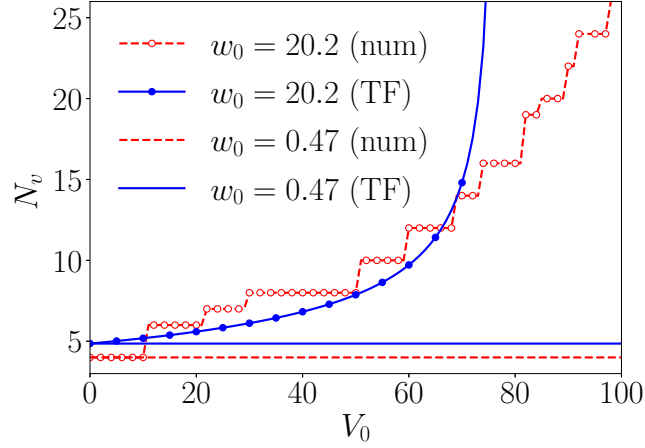}
\end{center}
\caption{Plot of the number of vortices ($N_v$) as a function of Gaussian potential depth ($V_0$) for $w_0=20.2$ and $w_0=0.47$. Red colour curves represent the numerically calculated values and blue colour curves denote the TF approximated values.}
\label{fig:vortcom}
\end{figure}

\begin{figure}[!ht]
\begin{center}
\includegraphics[width=0.5\linewidth]{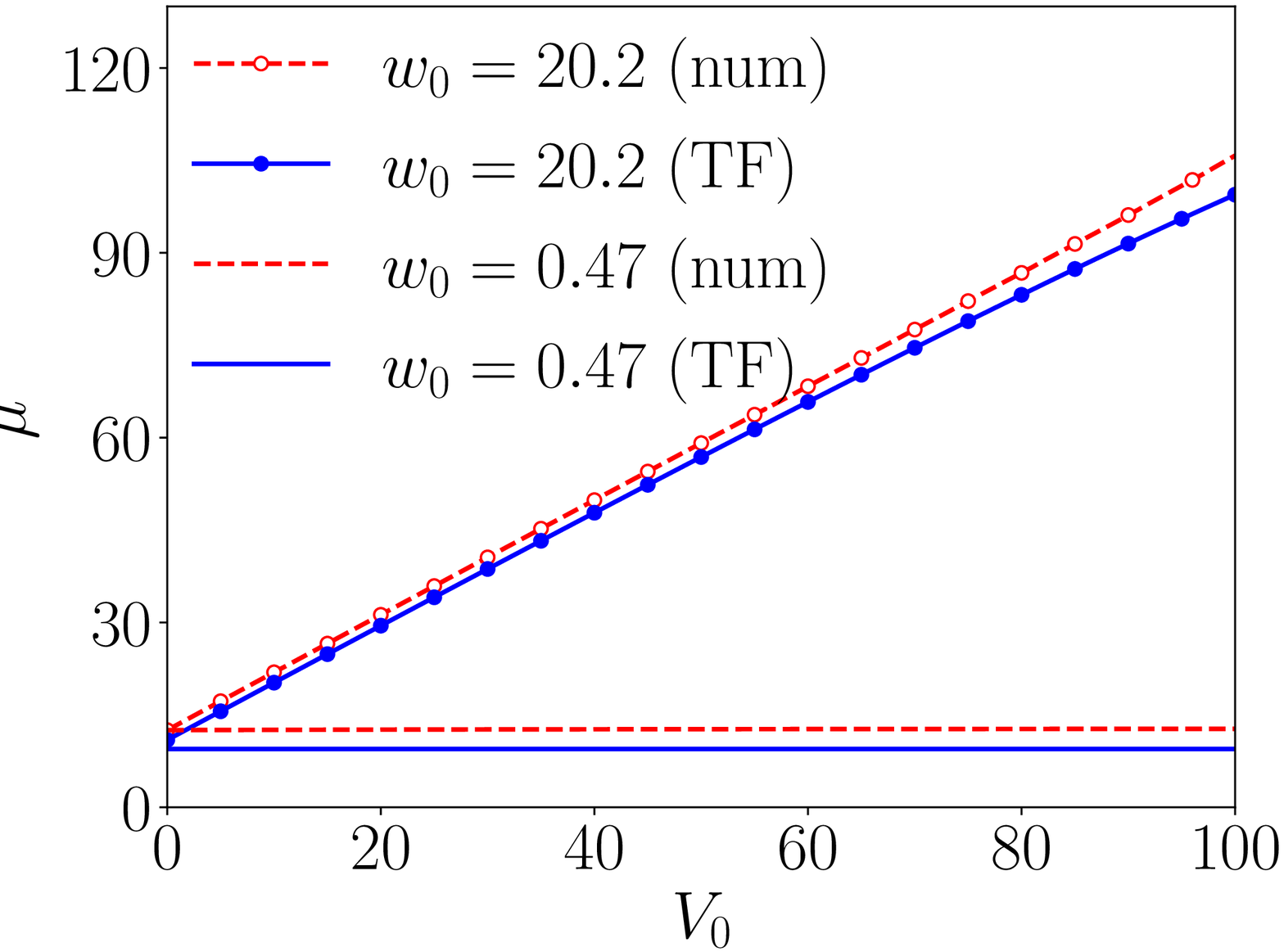}
\end{center}
\caption{Plot of the chemical potential ($\mu$) versus Gaussian potential depth ($V_0$) for $w_0=20.2$ and $w_0=0.47$. Red colour curves represent the numerically calculated values and blue colour curves denote the TF approximated values.}
\label{fig:chemcom}
\end{figure}

To understand the variation in the number of vortices with Gaussian potential depth ($V_0$), we calculate the rms radius ($r_{rms}$) for various values of $V_0$. We plot the calculated values of rms radius, number of vortices and chemical potential as a function of Gaussian potential depth for the two extreme values of $w_0$, say, $20.2$ and $0.47$ in Figs.~(\ref{fig:rmscom}), (\ref{fig:vortcom}) and (\ref{fig:chemcom}), respectively. %
In Fig.~\ref{fig:rmscom}, we plot $r_{rms}$ as a function of $V_0$ for largest and smallest $w_0$ values, that is, $w_0=20.2$ and $0.47$. The curves drawn in dashed line with red open circles are for the numerically calculated values, while the solid line with blue filled circles correspond to the TF approximated values. Fig.~\ref{fig:rmscom} shows that the rms radius, ($r_{rms}$), increases exponentially for $w_0=20.2$ while it remains constant for $w_0=0.47$. Consequently, the constancy of rms radius for $w_0=0.47$ irrespective of the values of Gaussian potential depth reveals the fact that the number of vortices within the condensate remains constant for all values of $V_0$ which can be seen from Fig.~\ref{fig:vortcom}. 

As far as the chemical potential is concerned, it linearly increases with Gaussian potential depth for $w_0=20.2$ as shown in Fig.~\ref{fig:chemcom} due to the heavier overlap of the condensate and the beam. While the curve stays flat for $w_0=0.47$ as depicted in Fig.~\ref{fig:chemcom} because of a minimal overlap between the condensate and the beam. In all three figures, namely Figs.~\ref{fig:rmscom}, \ref{fig:vortcom} and \ref{fig:chemcom}, it can be seen that the numerically calculated values corroborated very well with that of the TF approximation.

\subsection{Expectation value of angular momentum}

\begin{figure}[!ht]
\begin{center}
\includegraphics[width=0.5\linewidth]{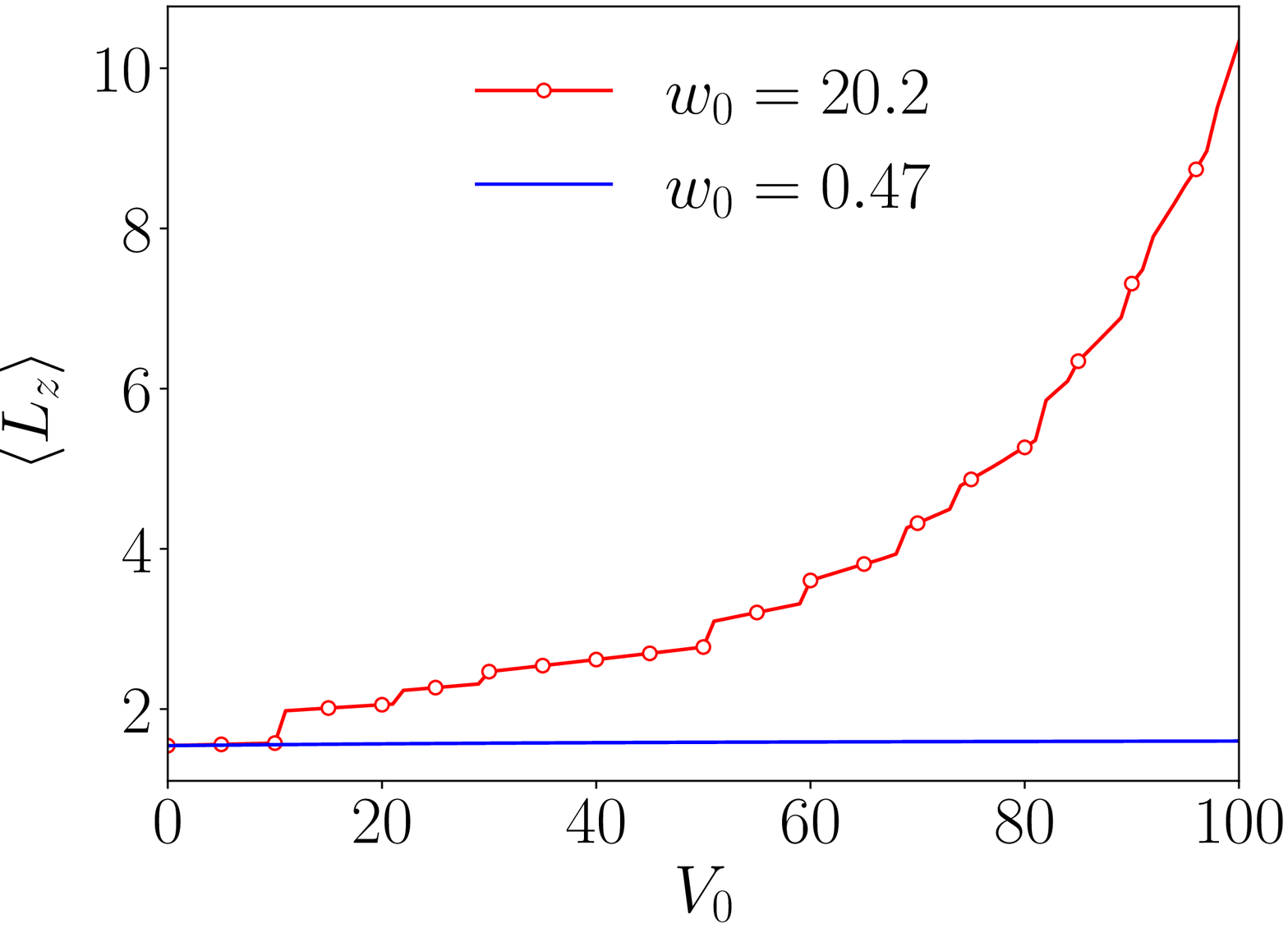}
\end{center}
\caption{Plot of the expectation value of angular momentum ($\langle L_z\rangle$) versus the Gaussian potential depth ($V_0$) for $w_0=20.2$ and $w_0=0.47$.}
\label{fig:angcom}
\end{figure}

To ascertain the birth of new vortices as a function of  waist and potential depth of the laser beam, we calculate the expectation values of angular momentum as a function of $V_0$ from the numerical solution. Fig.~\ref{fig:angcom}, depicts the variation of $\langle L_z\rangle$ with respect to $V_0$ for the two values of $w_0$, namely, $w_0 = 20.2$ and $0.47$. %
In Fig.~\ref{fig:angcom}, the expectation values of angular momentum ($\langle L_z\rangle$) seem to increase monotonically for $w_0=20.2$, which signifies the birth of new vortices within the condensate. On the other hand, understandably, for the thin beam due to the poor overlap of the beam with the condensate, these $\langle L_z\rangle$ values for $w_0=0.47$ remains almost constant, denoting no more new vortices generated within the condensate as shown in Fig.~\ref{fig:angcom}.

\subsection{Free energy}

Another significant quantity that demonstrates the creation of new vortices as the potential depth increases is the free energy $F$, which can be estimated as~\cite{Kato2011} 
\begin{align}\label{eq:free_energy}
F=E-\mu-\Omega \langle L_z \rangle,
\end{align}
where $E$ and $\mu$ correspond to the energy and chemical potential in the absence of rotation. Formation of vortices can be seen in terms of the free energy by computing the characteristic vortex nucleation time. Therefore, we plot a graph between the change in free energy ($\Delta F = F(t) - F(0)$) with respect to time %
\begin{figure}[!ht]
\begin{center}
\includegraphics[width=0.5\linewidth]{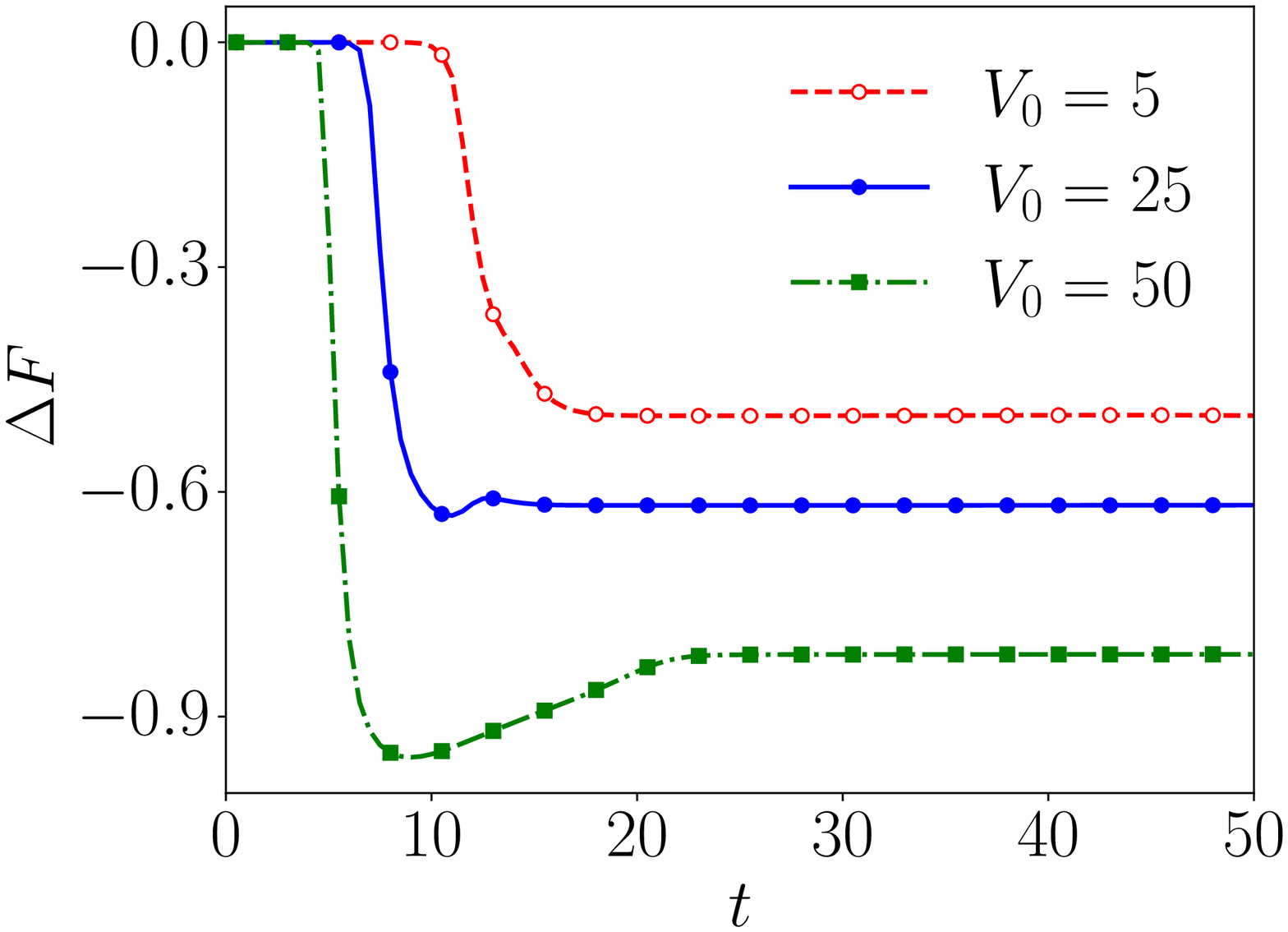}
\end{center}
\caption{The time evolution of the free energy for three different values of $V_0$  plotted as the difference between the instant free energy and the free energy at $t=0$ with $w_0=20.2$}
\label{fig:free-energy}
\end{figure}%
and observe that during the formation of a vortex, the free energy of the condensate rapidly decreases for some time and it stabilizes once the vortex reaches the equilibrium state. Also, from Fig.~\ref{fig:free-energy}, we observe that as the Gaussian potential depth increases, the characteristic nucleation time decreases. Moreover, the free energy fall-off is steeper for a higher depth of the Gaussian potential ($V_0$).

\subsection{Critical Rotation frequency}

Even though we keep the rotation frequency constant for our analysis, we are interested to know whether the considered value of rotational frequency is fair enough to suit the conditions for vortex formation. In addition to the above, we intend to study the relationship between the critical rotational frequency and the Gaussian potential depth for the formation of a single vortex. We calculate the critical rotational frequencies for different Gaussian potentials for two different waist values, viz $w_0=20.2$ and $0.47$, and also for the entire range of potential depth of interest. %
\begin{figure}[!ht]
\begin{center}
\includegraphics[width=0.5\linewidth]{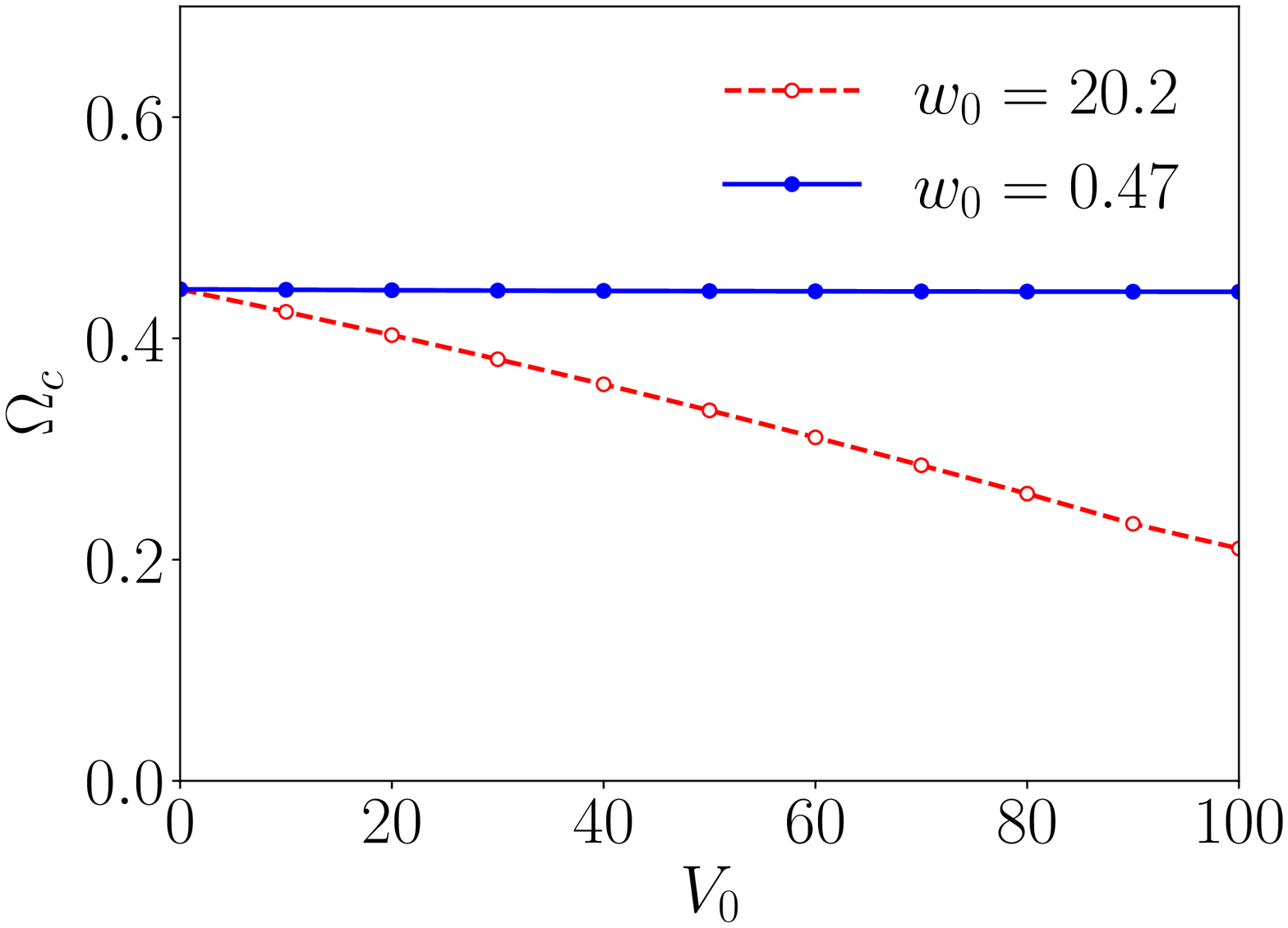}
\end{center}
\caption{Plot of critical rotational frequency ($\Omega_c$) as a function of Gaussian potential depth ($V_0$) for $w_0 = 20.2$ and $w_0 = 0.47$}
\label{fig:crit-ang}
\end{figure}%
In Fig.~\ref{fig:crit-ang}, we show the variation of critical rotational frequency as a function of $V_0$ for $w_0=20.2$ and $0.47$, respectively, with sufficiently long time evolution to ensure the equilibrium. It is clearly seen from Fig.~\ref{fig:crit-ang} that our choice of rotational frequency is well above the critical rotational frequency for the entire range, which bears a downtrend linear relationship with $V_0$ for $w_0=20.2$ and remains almost constant with $V_0$ when $w_0=0.47$.

\section{Summary and Conclusion}
\label{sec:conclu}
In this paper, we have studied the creation of vortices by triggering the rotating  Bose-Einstein condensates formed inside a toroidal trap with the trap parameters such as laser beam waist and Gaussian potential depth. By numerically solving the time-dependent Gross-Pitaevskii equation in two dimensions, we have shown that the vortex structure has undergone change when we vary the Gaussian potential depth. Also, we noticed a considerable increase in the number of vortices when the waist of the irradiated laser beam is in good agreement with the root mean square of the condensate. By computing the root mean square radius of the condensate, we have confirmed the variation in the number of vortices created as a function of the ratio between the root mean square radius of the condensate and the laser beam waist. The number of hidden vortices reaches the maximum value when the above ratio is close to $0.7$. In addition, we have found the variation in the number of vortices is abrupt for a deep Gaussian potential. In a nutshell, we conclude that the larger beam waist and a deep Gaussian potential have generated a lot of vortices. Also, we calculated the number of vortices using the Feynman rule with the Thomas Fermi approximation and compared them with the numerical results. Finally, we have observed that the critical rotation frequency decreases with an increase in the depth of Gaussian potential. Our results may throw some light on the applicability of rotating BECs towards the fabrication of quantum computers. Precisely, our work may be useful for the creation of phase vortex qubits, as discussed in ref.~\cite{Kapale2005}.

\subsection*{Acknowledgements}
The authors thank Dr. R. Kishor Kumar for helpful discussions. The work of P.M. is supported by CSIR under Grant No. 03(1422)/18/EMR-II, DST-SERB under Grant No. CRG/2019/004059, DST-FIST under Grant No. SR/FST/PSI-204/2015(C), MoE RUSA 2.0 (Physical Sciences) and DST-PURSE (Phase-II) Programmes. 


\begin{appendix}
\section{Expressions for inner and outer radii of an annular condensate}
For $w_0\gg r$, one can obtain the outer radius of the condensate ($R_2)$  by the following equation \cite{Aftalion2010}
\begin{align}
    \frac{8V_0}{15 w_0^8}R_{2}^{10}-\frac{V_0}{w_0^6}R_{2}^8+\frac{4V_0}{3 w_0^4}R_{2}^6+\frac{1}{4}\left(1-\Omega^2-\frac{4V_0}{w_0^2}\right)R_{2}^4-\frac{g}{\pi}=0
    \label{eqn:A1}
\end{align}
and the inner radius of the condensate $(R_1)$ can be obtained by the relation    
\begin{align}
    R_1\sim \frac{\xi}{\sqrt{\left[\left(1-\Omega^2-\frac{4V_0}{w_0^2}\right)R_2^2+
    \frac{4V_0}{w_0^4}R_2^4-\frac{8V_0}{3w_0^6}R_2^6+\frac{4V_0}{3w_0^8}R_2^8-\frac{8V_0}{15w_0^10}R_2^{10}\right]}}
\end{align}  
where $\xi$ is the quantum circulation of the condensate which can be obtained from the following equation 
\begin{align}
     \frac{g\Omega}{\pi \xi} \sim \frac{1}{2}\left(1-\Omega^2-\frac{4V_0}{w_0^2}\right)R_2^2\left[\ln(R_2^2)-2\right]+\frac{V_0}{w_0^4}R_2^4\left[2\ln(R_2^2)-3\right]-\frac{4V_0}{8w_0^6}R_2^6\left[3\ln(R_2^2)-4\right]   
\end{align}
For $w_0\ll r$, one can get the outer radius of the condensate by the following relation
\begin{align}
    R_2 \sim \left[\left(\frac{4g}{\pi}+2V_0w_0^2\right)\frac{1}{(1-\Omega^2)}\right]^{1/4} \label{eqn:A4}
\end{align}
and hence the inner radius of the condensate can be expressed as a quadratic equation in $(R_1^2)$ as follows:
\begin{align}
    \frac{4V_0}{w_0^2}R_1^4+[(1-\Omega^2)R_2^2-2V_0]R_1^2-\xi^2=0 \label{eqn:A5}
\end{align}
with the quantum circulation
\begin{align}
\xi \sim \frac{2g\Omega}{\pi(1-\Omega^2)R_2^2[\ln(R_2^2)-2]+8.59V_0}.
\end{align}
\end{appendix}
\end{document}